\documentclass[10pt, conference]{IEEEtran}
\IEEEoverridecommandlockouts
\usepackage{cite}
\usepackage{amsmath,amssymb,amsfonts}
\usepackage{graphicx,setspace}
\usepackage{textcomp}
\usepackage{xcolor}
\PassOptionsToPackage{hyphens}{url}\usepackage{hyperref}
\usepackage{algorithm,algpseudocode}
\usepackage{textcomp}
\usepackage{array}
\usepackage{tabularx}
\usepackage{booktabs}
\usepackage{mathtools}
\usepackage{comment}
\usepackage{amsthm}
\usepackage{url}

\usepackage{epstopdf}

\usepackage{subfigure}
\usepackage{makecell}
\usepackage{comment}
\usepackage{marvosym}
\usepackage{booktabs}

\newcommand\blfootnote[1]{%
	\begingroup
	\renewcommand\thefootnote{}\footnote{#1}%
	\addtocounter{footnote}{-1}%
	\endgroup
}

\theoremstyle{definition}

\allowdisplaybreaks

\begin{document}
\title{HAPS-ITS: Enabling Future ITS Services in Trans-Continental Highways
}

\author{\IEEEauthorblockN{Wael Jaafar and Halim Yanikomeroglu}
}

\maketitle

\begin{abstract}
With the advent of rapid globalization and the inter-border supply chain network, the reliability and efficiency of transportation 
systems have become even more critical. Indeed, trans-continental highways need particular attention due to their important role in sustaining globalization. In this context, intelligent transportation systems (ITS) can actively enhance the safety, mobility, productivity, and comfort of trans-continental highways. However, ITS efficiency depends greatly on the roads where they are deployed, on the availability of power and connectivity, and on the integration of future connected and autonomous vehicles. To this end, high altitude platform station (HAPS) systems, due to their mobility, sustainability, payload capacity, and communication/caching/computing capabilities, are seen as a key enabler of future ITS services for trans-continental highways; this paradigm is referred to as HAPS-ITS. The latter is envisioned as an active component of ITS systems to support a plethora of transportation applications, such as traffic  monitoring, accident reporting, and  platooning. This paper discusses how HAPS systems can enable advanced ITS services for trans-continental highways, presenting the main requirements of HAPS-ITS and a detailed case study of the Trans-Sahara highway.




\end{abstract}

\section{Introduction}

There was a time when traveling within and across countries meant riding for months in horse-drawn wagons or camel-led caravans. With the development of industrialized transportation systems, all of this has changed. Given the widespread availability of automobiles and train travel in the 20th century, a whole new era began, involving
overland roads, canals, bridges, and railways. 
These routes changed our way of life, allowing people and goods to move between cities and across areas previously considered uninhabitable. In the process, the economical, social, cultural, and territorial aspects of globalization has accelerated. 
One of its major vectors is the development of trans-continental highways. The latter are primary arteries that connect cities within a single country, such as Trans-Canada highway (7,821 km), Trans-Siberian highway (11,000 km), and Australian Highway 1 (14,500 km), or many cities across different countries, e.g., Trans-African highway network (56,683 km).
\blfootnote{\textcolor{red}{This work has been accepted for publication in IEEE Communications Magazine. Copyright may be transferred without notice, after which this version may no longer be accessible.}}

Another factor driving the development of highway infrastructure is the global market of cellular vehicle-to-everything for intelligent transportation systems (ITS), which is expected to exceed \$1 trillion US circa 2030.
This will be driven by the growing integration of advanced technologies to manage vehicular traffic in densely populated mega-cites and road infrastructures creaking under the strain of rapid urbanization and growing number of vehicles. Also, since road transportation accounts for 27\% in global CO$_2$ emissions, the use of ITS tools is advocated to reduce transportation carbon footprint \cite{Bojin2012}. Indeed, ITS incorporates technologies such as Internet-of-things (IoT), radar, data processing, dissemination/transmission, and intelligent control, to improve safety, mobility, productivity, and comfort. 

Unlike urban areas where ITS focuses on productivity, on highways the primary focus is on safety and reducing mortality rates \cite{Cody2016}. The issue of fatalities on trans-continental highways is accentuated by the fact that several road segments may be totally isolated. Although road infrastructures have been benefiting from ITS for decades, e.g., path finder, dedicated short-range communications (DSRC), and variable messages signs (VMS), a new generation of ITS technologies, such as connected and autonomous vehicles (CAVs), are creating novel applications that demand smart infrastructure involvement.  

The lack of connectivity along trans-continental highways is a major limiting factor for ITS goals. Indeed, vehicles rely only on their own intelligence and occasional vehicle-to-vehicle (V2V) communications to improve road safety, establish platooning, and reduce their carbon footprint, while services that require remote control or cloud services cannot be realized for lack of efficient vehicle-to-infrastructure (V2I) communications.

\begin{figure*}[t]
	\centering
	\includegraphics[width=1\linewidth]{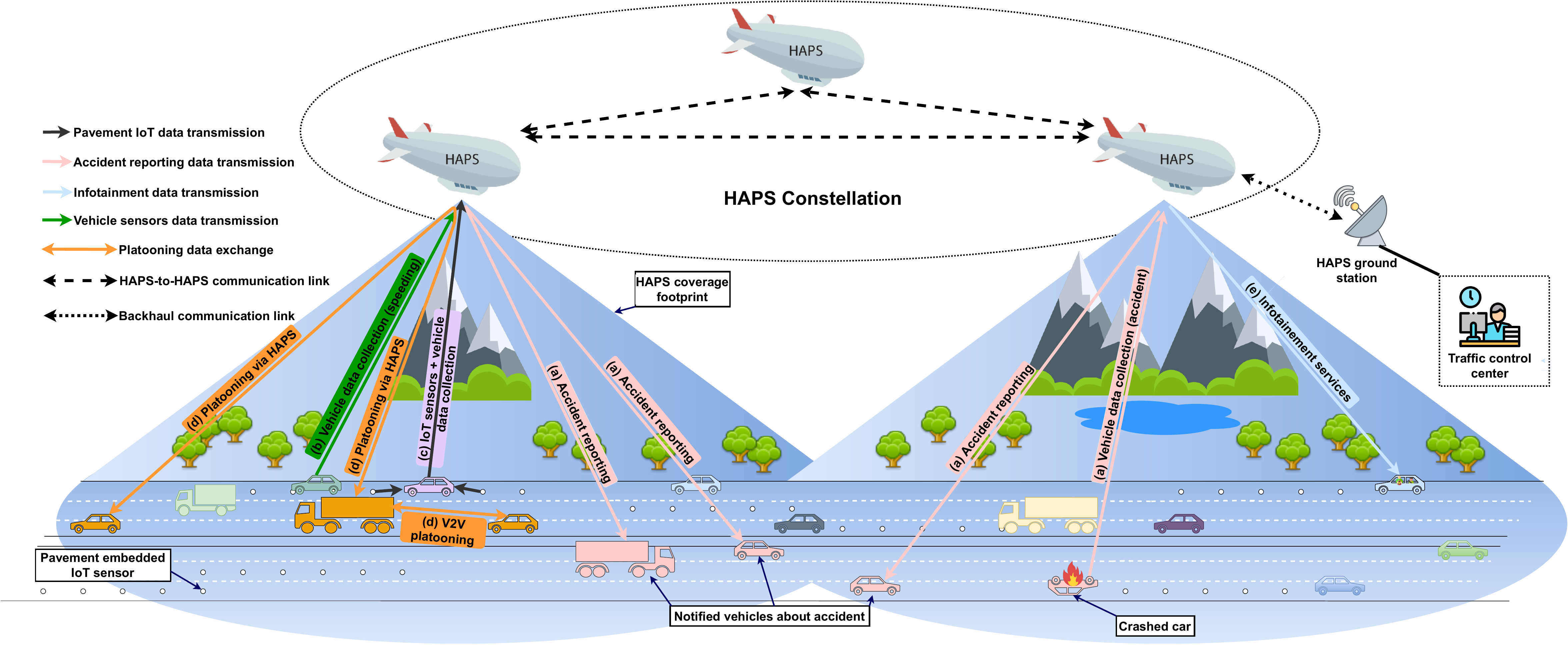}
	\caption{Example of data flows for HAPS-supported ITS services.} 
	\label{fig:sys}
\end{figure*}

Although satellites can be leveraged for ITS services in any area, their use is limited to delay-tolerant services, while critical ITS applications cannot be handled.
To tackle this issue, the use of a constellation of high altitude platform stations (HAPS) is advocated. In this paper, we envision HAPS as the main enabler of future ITS services in trans-continental highways. Indeed, a HAPS node is a wireless network node that operates at a typical altitude of 20 km. Due to recent innovations in autonomous avionics, array antennas, battery, and solar energy, HAPS systems are emerging as a principal component of next-generation networks \cite{kurt2020vision}. In industry, several HAPS start-ups are leading the way towards high-speed connectivity from the stratosphere, including HAPSMobile, 
Thales Alenia Space, 
and Stratospheric Platforms Limited.
As a super macro base station (BS) \cite{alam2020high}, a HAPS node is expected to provide wireless and Internet connectivity in a wide area up to 500 km in radius (ITU-R F1500), thus enabling several ITS applications like traffic monitoring, accidents reporting, platooning, and sensors data collection. Moreover, we propose HAPS as an aerial data-center capable of processing big data for ITS services such as road
traffic accident analysis, traffic prediction and route planning, and fleet management and control \cite{Zhu2019}.
Similarly, supported by a high storage payload, HAPS caching can be leveraged to provide infotainment services to passengers, e.g., video streaming and gaming. 
Since communication, computing, and caching resources may not be sufficient in a single HAPS, we envision the use of multiple HAPS nodes with high HAPS-to-HAPS (H2H) data rate links to handle ITS services and improve reliability in case of failures.

\section{HAPS-ITS for Enhanced ITS Services in Trans-Continental Highways}
Given the limited government budgets, innovation is needed to get the best from existing infrastructure assets. Also, with the proliferation of CAVs, transportation networks need to be ready with adequate ITS technologies to support CAV services.

\subsection{ITS Challenges in Trans-Continental Highways}


Although new ITS functionalities are attractive for making highways safer, more efficient, and eco-friendly, a number of challenges need to be addressed. First, trans-continental highways suffer from spotty cellular coverage due to harsh terrain conditions, e.g., mountain areas, and from a lack of economic interest by service providers.
Consequently, critical ITS services, such as accident reporting, are inoperable in uncovered areas. Also, real-time infrastructure condition and traffic monitoring cannot be realized in the absence of V2I communications.
Since CAV functions are mainly executed by an on-board computer, the latter may potentially fail. 
In such a situation, the vehicle becomes either ``blind'' and relies fully on the driver, or leverages V2V and V2I links to offload its tasks to another vehicle or a roadside unit (RSU). Nevertheless, this alternative requires the presence of close-by vehicles and/or cellular coverage, which may not be available. In addition, conventional V2V and V2I links may experience security issues, due to malicious attacks injecting erroneous data to misguide CAVs.
Finally, long trips along trans-continental highways may be stressful, due to the absence of infotainment services, such as accurate navigation, news updates, and entertainment.  
Hence, there is an urgent need for ITS enabling solutions in trans-continental highways.

To tackle the aforementioned challenges, we advocate the use of a HAPS constellation as a reliable provider of connectivity, caching, computing, and imaging power for ITS services.
Several {CAV} applications can be enabled through HAPS-ITS, thus providing improved travel safety,
better productivity, 
and greater mobility and comfort for passengers 

In the remaining, HAPS-ITS refers to unmanned airships equipped with components that enable ITS services. This is motivated by the numerous advantages of unmanned airships over unmanned aircraft, including quasi-stationarity, large payload (up to 2000 kg), and extended mission times (up to 5 years).

\subsection{HAPS-ITS for Improved Safety}
As passenger safety is the primary concern on highways, the capacity to respond to accidents in real-time increases the chance of saving lives. Indeed, 
passengers need to feel safe and supported in case of a hazard.

\subsubsection{HAPS as an accident reporting agent} 
Empowered by massive multiple-input-multiple-output (m-MIMO) antennas and hybrid radio frequency (RF) and free-space optical (FSO) backhauling \cite{Alzenad2018},
HAPS-ITS can support emergency calls from isolated areas and thus allows an immediate response. Also, by adopting the new radio vehicle-to-everything (NR-V2X) communication protocol, it can guarantee V2I-supported services, such as informing vehicles of upcoming hazards, 
monitoring their speed, and the status of roads. If an accident occurs, the HAPS system can identify it by analyzing the vehicle's data, as shown in use-case (a) of Fig. \ref{fig:sys}. To assess the situation rapidly, first responders would rely on calls incoming from victims or witnesses, combined to the captured images by the high-resolution cameras on-board the HAPS nodes. 
In fact, camera technology has improved so greatly that 
LiDAR cameras can constitute objects' pictures from 45 km away \cite{Li2020}. 

\subsubsection{HAPS as a surveillance agent} 
In politically/socially unstable regions, roads crossing
borders may 
present opportunities for criminal activities.
By deploying HAPS-ITS over them, on-board cameras can monitor traffic, identify illegal activities, and respond rapidly to them. 
Also, data collected from CAVs through the NR HAPS-to-vehicle (H2V) links can be used to monitor vehicle speed and/or issue tickets, as illustrated in example (b) of Fig. \ref{fig:sys}. The integrity of this data can be validated by the captured video, and thus bypass any malicious data alteration. For efficient surveillance, artificial intelligence algorithms such as ``You Only Learn One Representation'' (YOLOR) for object detection and tracking, can be called in to analyze live data, detect threats, and trigger timely responses. 

\subsubsection{HAPS as an aerial roadside unit} In the future, intelligent road equipment will be intensely deployed in highways. Intelligent roads rely on IoT sensors embedded within the pavement tapes or markers. These sensors can operate with minimum maintenance using partial energy autonomy \cite{Verma2021}. Their role includes supporting road maintenance, dynamic speed limiting for different types of vehicles, and providing information to CAVs about road conditions, weather, and traffic. To do so, they adopt communication protocols such as Wi-Fi, bluetooth, and long-range wide-area network (LoRaWAN). Along isolated highway segments, these sensors can rely on solar energy to operate, while connectivity to the control center is provided through HAPS. 
Due to the limited power of sensors and to the high path loss, transmitting IoT data directly to HAPS may not be feasible. Alternatively, sensors can exploit passing-by CAVs as relays to forward data to the HAPS node, as shown in use-case (c) of Fig. \ref{fig:sys}. Indeed, communication between sensors and CAVs can be established within a few hundred meters, e.g., using Wi-Fi or LoRaWAN, and last for a few seconds, enough to transmit IoT packets with minimal power.
Moreover, V2V communications are handled by on-board transmit/receive hardware. If {it} fails, the vehicle cannot react to other vehicles' messages. This issue can be bypassed by forwarding V2V messages through HAPS. Despite the additional round-trip delay, to and from the HAPS node, ranging between 0.13 and 0.33 ms \cite{alam2020high}, the overall delay is
still small to trigger timely actions when needed. {To ensure this function, HAPS systems need to support NR-V2X communication protocols (3GPP Release-17)}.

\begin{figure}[t]
	\centering
	\includegraphics[width=0.9\linewidth]{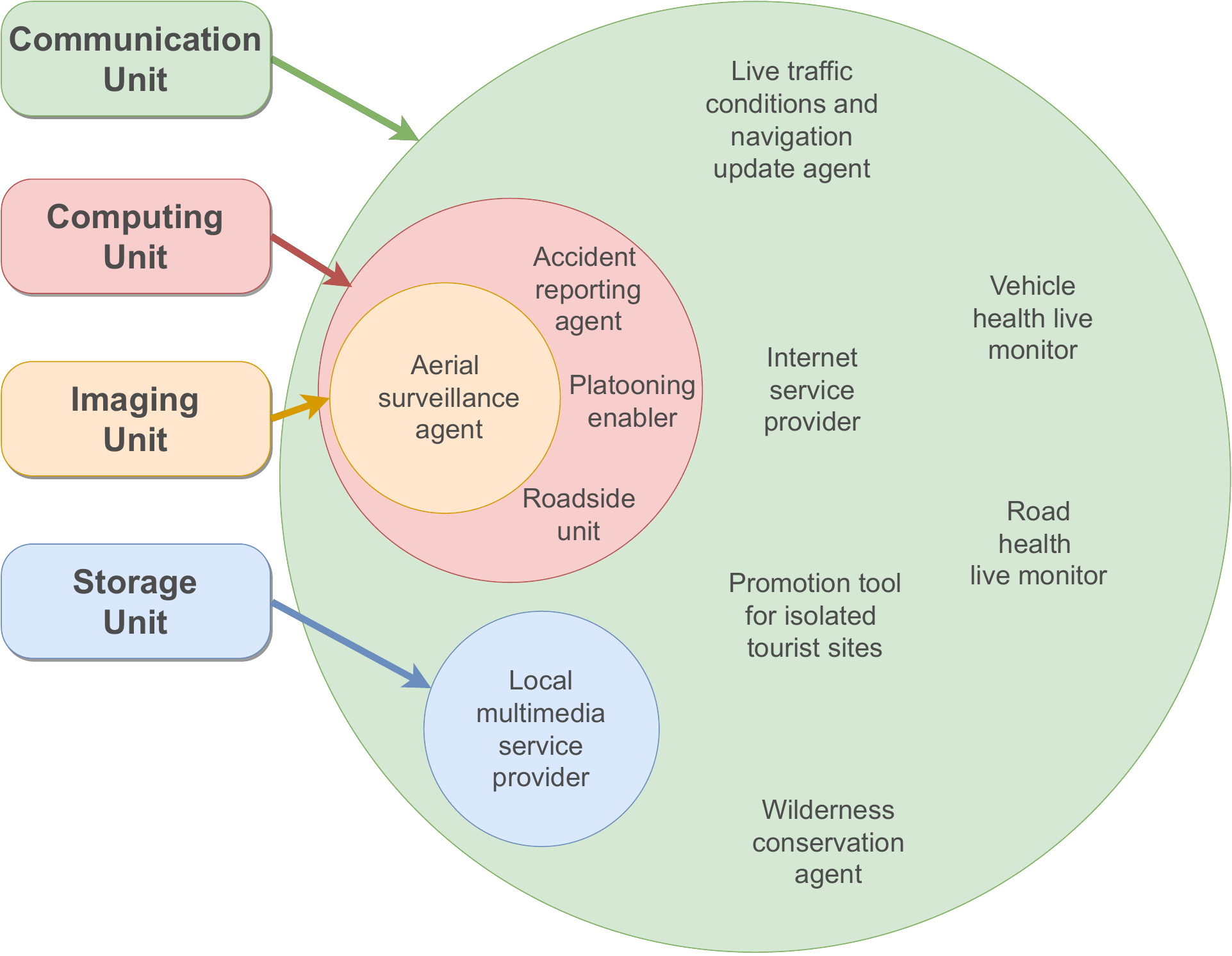}
	\caption{Relation between the targeted trans-continental highway services and their main enabling HAPS-ITS components (a service placed in a circle within another circle means that both units are the main drivers of that service).
	}
	\label{fig:HAPS_ITS}
\end{figure}

\subsection{HAPS-ITS for Better Productivity }
As 85\% of freight traffic travels by road, governments invested billions of dollars in highway extensions. 
However, highway networks require {patrolling and maintenance, often carried out by} limited funding. {To reduce costs, remote monitoring of vehicles/roads, and efficiently slowing or rerouting traffic can be leveraged by ITS services. The latter}
require V2I communications, {which may not be available.} 
HAPS-ITS can be deployed as an alternative to support V2I and provides the necessary computation power for such services. As illustrated in example (c) of Fig. \ref{fig:sys}, through the collection and analysis of the pavement sensors data, HAPS-ITS can monitor highway segments, identify critical spots, and alert the control center.


Also,
platooning has been proven efficient to coordinate traffic, reduce congestion, and cut carbon footprint. 
It is even necessary when traveling along trans-continental highways for thousands of kilometers. Typically, it relies on V2V to exchange messages in the convoy, such as vehicle accelerating/decelerating, braking, etc. However, if a communication failure occurs, the convoy may lose formation and platooning benefits. V2I connectivity through HAPS-ITS not only allows a platoon to bypass failed V2V links, but also to leverage platooning for vehicles not yet in V2V range, as demonstrated in use-case (d) of Fig. \ref{fig:sys}. Indeed, with the HAPS-ITS's wide view of the highway, it can incentivize platooning and organize convoys.





\subsection{HAPS-ITS for Higher Mobility}
Providing reliable and fast Internet coverage along highways has several mobility advantages. 
First, it increases the visibility and popularity of isolated and unknown tourist sites.
Thus, local economy is positively stimulated, new businesses launched, and
number of citizens permanently living in the area increased.
In contrast to terrestrial BSs, a HAPS network provides connectivity without affecting the wilderness and beauty of the natural environment. This is important since several strict regulations have been put in place to preserve important sites and parks from visual pollution \cite{commission2017}.   
Finally, as highway traffic increases, HAPS-ITS processing power enables efficient traffic management 
through, for instance, live updates of VMS and CAVs' navigation systems.

\subsection{HAPS-ITS for Enhanced Comfort}
Comfort of passengers in long trips along trans-continental highways is an important issue.
Whether a vehicle is autonomous or connected, passengers appreciate {reaching quickly} their destination.
{This can be enabled through HAPS-ITS by transmitting real-time traffic condition updates and rerouting information.} 
If the vehicle is autonomous, neither the driver nor the passengers need to watch the road, as driving is monitored by the vehicle's computer and supported by the HAPS-ITS' V2I information. In such a case, passengers can use the Internet service to surf, play games, or watch videos, as shown in example (e) of Fig \ref{fig:sys}. To reduce data traffic through the HAPS-ITS backhaul link and improve the traveling experience, the former can use its high caching/processing {power} and H2H links to act as a local multimedia server and provide vehicles with requested applications and services.

In Fig. \ref{fig:HAPS_ITS}, we summarize the relation between the targeted ITS services and the HAPS-ITS components.





\section{HAPS-ITS Requirements}

Since different levels of technology may be integrated into different CAVs, we address the characteristics of HAPS-ITS required to support the following CAV technology levels \cite{Liu2021}:
\begin{itemize}
    \item \textit{Level 1:} CAVs equipped with driver assistance features, e.g., adaptive cruise control and lane keep assistance. 
    \item \textit{Level 2:} Partially automated CAVs, which assist in controlling speed, for instance by maintaining distances from other vehicles in stop-and-go traffic and steering to center the vehicle within the lane. 
    \item \textit{Level 3:} Conditional automated CAVs are an upgrade of level 2 CAVs with autonomous operation under ideal conditions and limitations only, e.g., limited-access divided highways at a given speed. A human driver is still needed to take over if driving conditions fall below ideal.
    \item \textit{Level 4:} Highly automated CAVs can operate by themselves, without human drivers, but they are restricted to known cases only, e.g., an autonomous bus itinerary.
    \item \textit{Level 5:} Fully automated CAVs are true driveless vehicles, capable of monitoring the environment and maneuvering through any road conditions. 
\end{itemize}

As shown in Fig. \ref{fig:HAPS_ITS}, the on-board components involved in HAPS-ITS services are the communication unit, the storage unit, the computing unit, and the imaging unit. We describe them as follows.

\subsection{Communication Unit}
The \textit{communication unit} is responsible for supporting different types of communication links, namely H2H, H2V, vehicle-to-HAPS (V2H), and HAPS-to-gateway (H2G). High throughput H2H and H2G can be supported through millimeter wave (mmWave) or FSO technologies \cite{Taori2015,Alzenad2018}. For H2V/V2H, the throughput requirement may vary in accordance with the CAV level. Indeed, we estimate the CAV data generation rate to be \cite{counterpoint2019,Liu2021}:
\begin{itemize}
    \item \textit{Level 1:} Between 20.05 megabyte per second (MBps) and 40.5 MBps
    \item \textit{Level 2:} Between 120.07 MBps and 240.7 MBps
    \item \textit{Level 3:} Between 150.09 MBps and 350.9 MBps
    \item \textit{Level 4:} Between 160.13 MBps and 421.3 MBps
    \item \textit{Level 5:} Between 181.7 MBps and 561.7 MBps
\end{itemize}
\noindent
Subsequently, only a part of this data needs to be uploaded to the HAPS-ITS when traveling. Given that most of the traffic data processing is executed locally by the CAV's on-board computer and due to several security and privacy concerns, we can realistically assume that only 10\% of this data is relevant to the HAPS-ITS. Hence, this would be equivalent to transmitting at data rates below 32, 192, 281, 337, and 450 megabits per second (Mbps) for CAV levels 1, 2, 3, 4, and 5, respectively.
This may be achieved by HAPS-mounted large phased-array antenna, i.e., m-MIMO, which would provide connectivity through narrow beams and using MIMO antennas at the CAVs.
HAPS-ITS can provide several services, including infotainment, fleet management and maintenance, as well as operating systems and applications. 
Finally, safety services require a vehicle response time below 200 ms (response time of a professional driver). The HAPS-ITS can guarantee this requirement due to its short communication delay (below 0.33 ms) \cite{alam2020high}. 
However, achieving a high H2V throughput is challenging as current wireless technologies, namely LTE, mmWave, and m-MIMO, realize cell data rates under 500 Mbps at distances up to 40 km, which may handle the traffic of low level CAVs or of a small number of high level CAVs simultaneously. For denser traffic along trans-continental highways, the H2V link may rapidly become the HAPS-ITS bottleneck, which would limit the efficiency of the ITS services.
Finally, communication functions require a substantial number of components, such as antennas, transponders, low-noise power amplifiers, frequency
converters, and filters. Since one of the HAPS-ITS roles is similar to a BS, more active components may be needed, which would increase the power consumption and require more payload space in the HAPS nodes \cite{kurt2020vision}.

\subsection{Storage Unit} It is expected that CAVs will be equipped with \textit{storage units} between 2 terabytes (TB) and 11 TB, depending on the automation level \cite{counterpoint2019}. Hence, we expect the HAPS-ITS storage to range between 10 TB and 100 TB in order to handle different CAV automation levels for trans-continental highways, besides providing other services such as infotainment and fleet management. For instance, a level 1 vehicle would generate data at a rate of 145.8 GB every hour, while a level 5 vehicle's data would occupy 
up to 2 TB per hour. Equipping a HAPS-ITS with a large storage would ensure that it could occasionally handle operations for several vehicles, e.g., accident reporting and platooning, in the event that the latter's storage and/or processing equipment fails. 



\subsection{Computing Unit} In CAVs, processing units collect radar, and/or LiDAR and/or camera data and analyze them to provide ITS functions that range from driving assistance (level 1) to full driving automation (level 5). Thus, the computing requirements vary depending on the CAV level. For levels 1-2, anywhere from a few dozen to a few hundred million operations per second (MOPS) would be required to provide minimum safety assistance to drivers and infotainment services. These computing requirements can be met, for instance, with Intel processors having 32-bit microcontroller units. 
As CAV level 3-5 involve features that approach full autonomous driving, more powerful computing components are required. In this matter, NVIDIA has been leading the industry with its NVIDIA DRIVE AGX autonomous vehicle solution. NVIDIA DRIVE AGX Pegasus is the most powerful system-on-chip (SoC)
capable of executing 320 trillion operations per second (TOPS), thus outperforming by 10 times the NVIDIA Jetson AGX Xavier solution\footnote{1 TOPS = 10$^6$ MOPS.}.
The computing unit of a HAPS-ITS must occasionally handle intensive tasks for a few CAVs at the same time. In other words, the HAPS-ITS can be equipped with one NVIDIA Jetson AGX Xavier to handle level 1-2 operations. However, at least two NVIDIA DRIVE AGX Pegasus SoCs would be required to support level 3-5 CAVs in trans-continental highways.




\subsection{Imaging Unit}
\textit{High-resolution cameras} are an interesting add-on to the HAPS-ITS in order to monitor critical trans-continental highway segments, rapidly assess incidents, and handle inaccuracies in collected IoT data.
Several technologies can be used for imaging, including optical cameras, radars, and LiDAR, which are now able to distinguish objects at very large distances \cite{Li2020}. These components can be installed in the HAPS-ITS to increase surveillance efficiency. 

\section{A Case Study}

In this section, we simulate a HAPS-ITS deployment to provide ITS services for the Trans-Sahara highway, from Algiers to Lagos (4,504 km). 
The choice of this highway is motivated by the fact that it crosses three countries, namely, Algeria, Niger, and Nigeria, and about 70\% of its route runs through the Sahara. For the sake of simplicity, results will be shown for the Trans-Sahara highway portion of Algeria only.

\begin{figure}[t]
	\centering
	\includegraphics[width=0.97\linewidth]{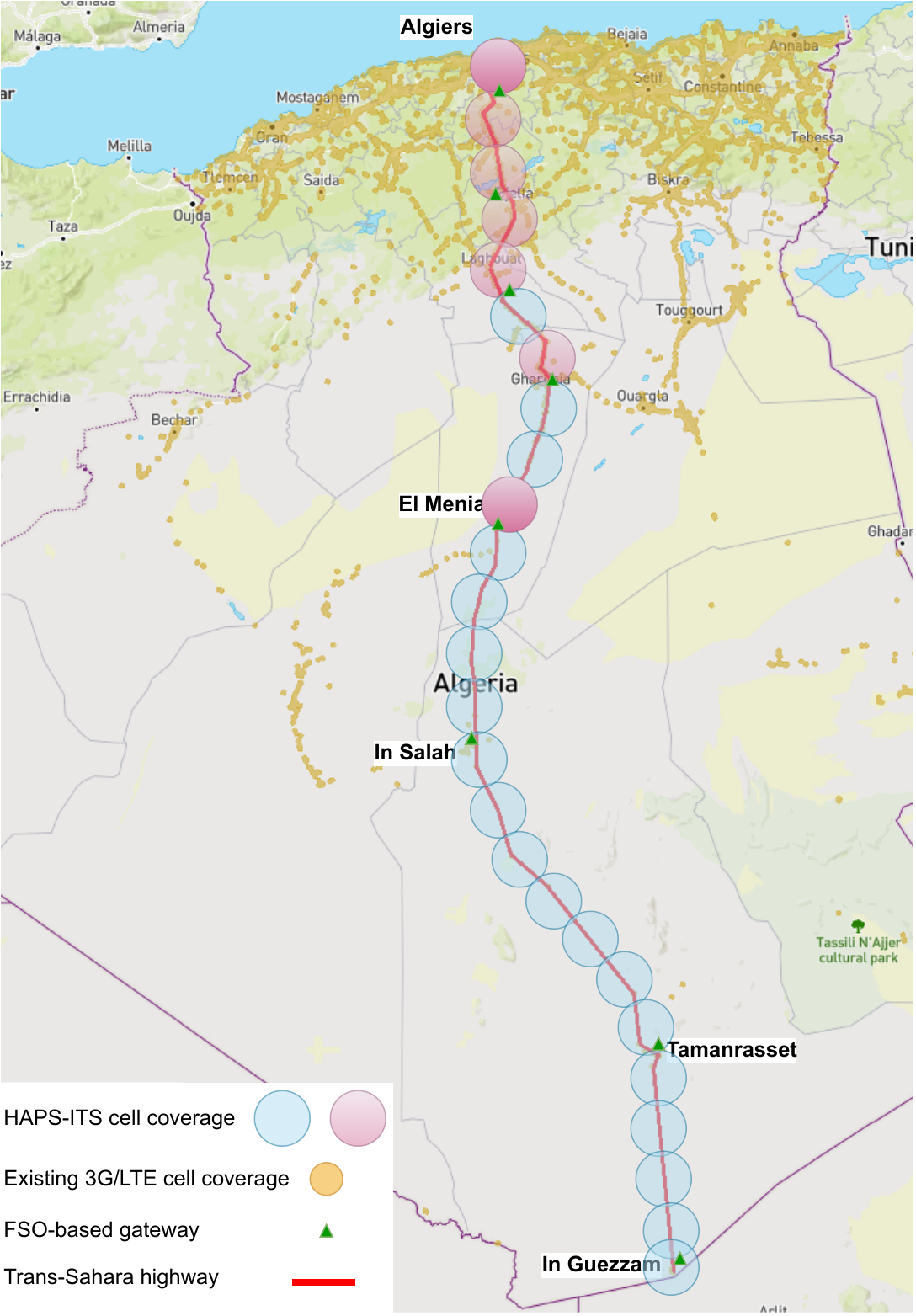}
	\caption{HAPS-ITS deployment along Algeria's Trans-Sahara highway (Blue+red cells to support ITS services without the terrestrial network; Blue cells only to support ITS services in collaboration with the terrestrial network).}
	\label{fig:Algeria}
\end{figure}

Assuming the HAPS-ITS nodes are deployed with the typical coverage footprint of 40 km in radius \cite{kurt2020vision}, approximately 59 HAPS-ITS nodes are required to provision ITS services for the Trans-Sahara highway, among which 26 are deployed in Algeria as shown in Fig. \ref{fig:Algeria}.
This number can be reduced if the existing cellular network along the highway supports ITS services. Indeed, 48 (resp. 19) HAPS-ITS nodes could fill the coverage gaps along isolated segments of the highway (resp. of the Algeria highway portion) if the existing 3G/LTE networks{\footnote{\url{www.gsma.com/coverage}}} provision ITS services, as illustrated in Fig. \ref{fig:Algeria}.

\begin{figure}[t]
	\centering
	\includegraphics[width=0.97\linewidth]{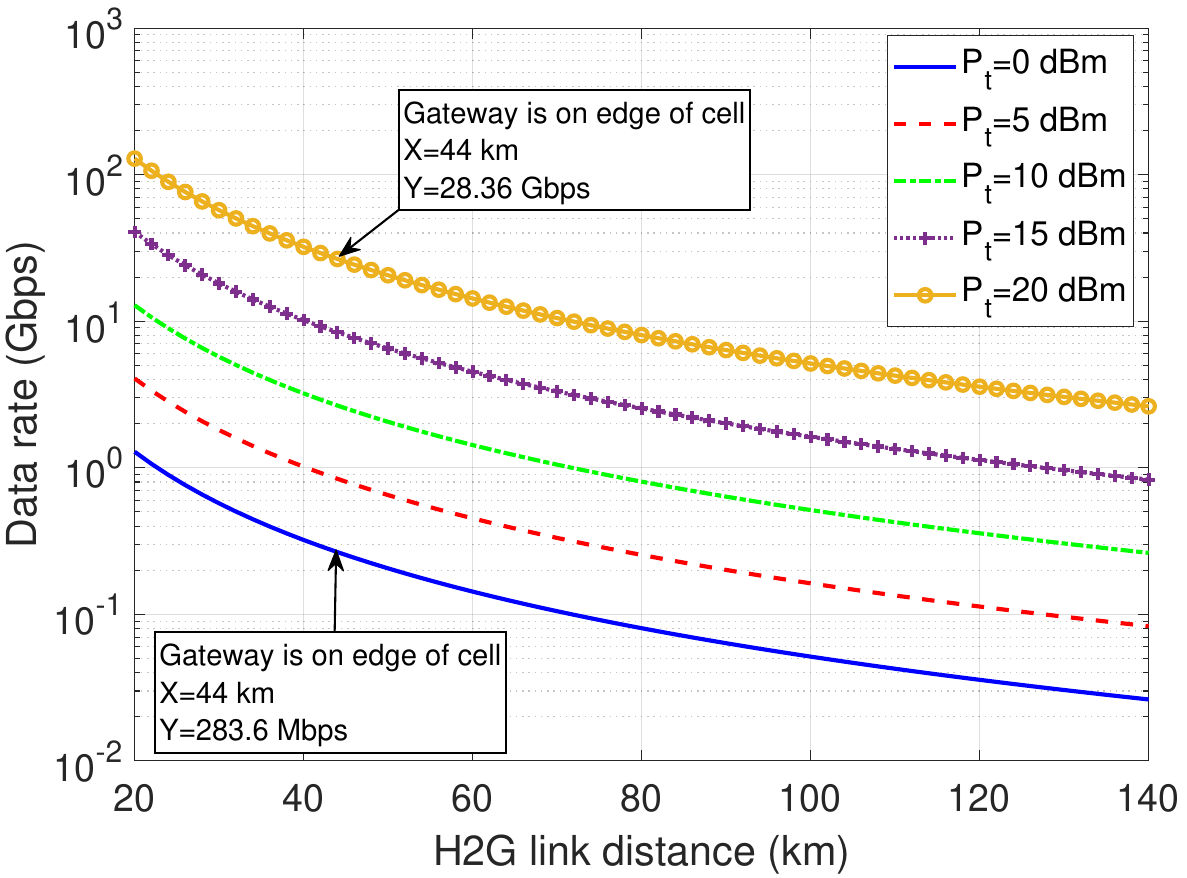}
	\caption{Data rate of FSO-based H2G link.}
	\label{fig:FSO1}
\end{figure}
\begin{figure}[t]
	\centering
	\includegraphics[width=0.96\linewidth]{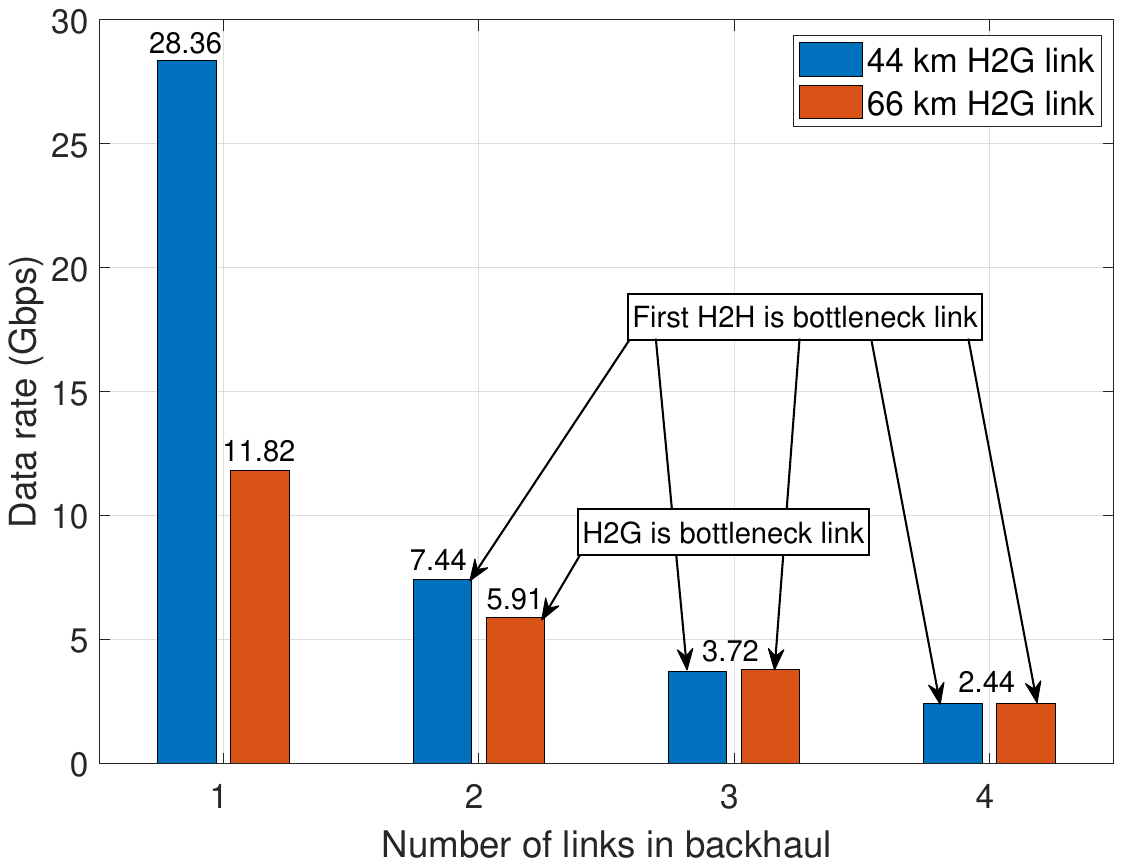}
	\caption{Data rate of FSO-based multi-hop backhaul ($P_t=20$ dBm).}
	\label{fig:FSO2}
\end{figure}

\begin{figure}[t]
	\centering
	\includegraphics[width=0.96\linewidth]{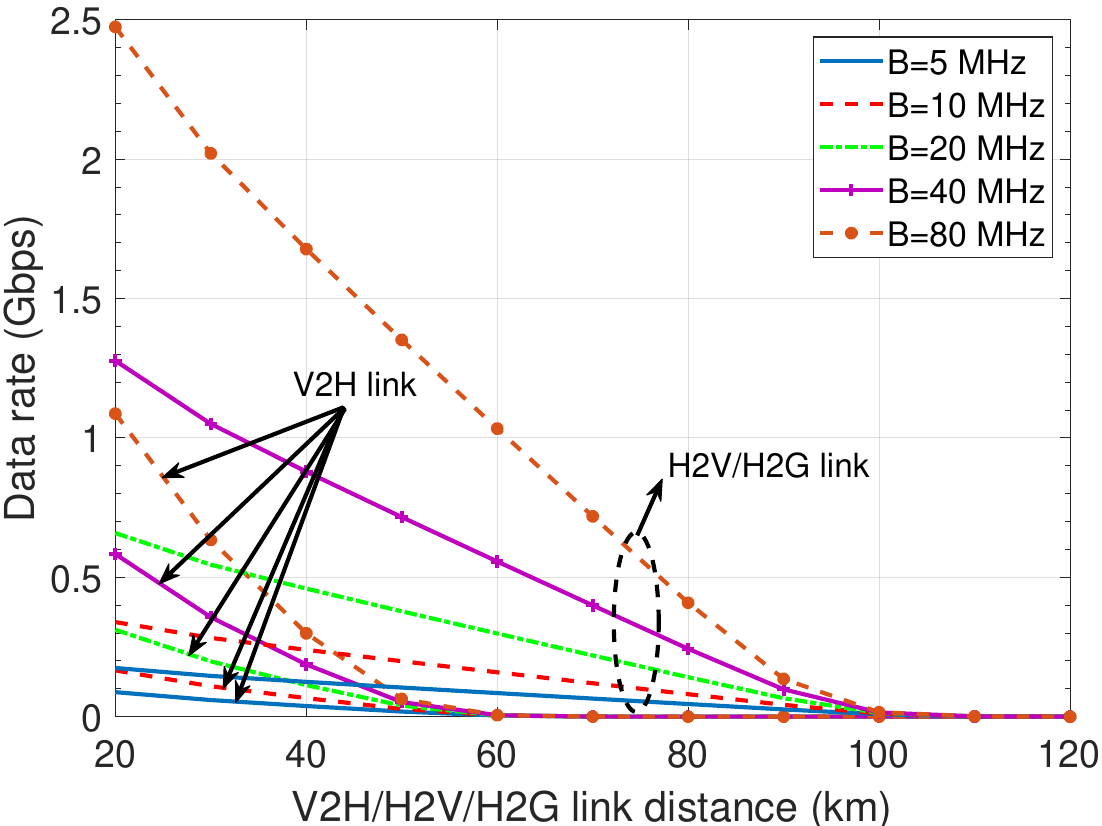}
	\caption{Data rate of mmWave V2H/H2V/H2G links.}
	\label{fig:mmWave}
\end{figure}

To support HAPS-ITS operations, a number of gateways are deployed. 
In Fig. \ref{fig:Algeria}, we see the deployment plan of FSO-based gateways (green triangles) in Algeria.
Practically, only 8 gateways can be deployed in cities to provide backhaul links to the HAPS-ITS constellation. Thus, several HAPS-ITS nodes along the El Menia-In Guezzam corridor rely on H2H links for backhauling. 

Fig. \ref{fig:FSO1} shows the data rate performance of the H2G FSO link as a function of its distance for different transmit power values $P_t$. Assumptions, parameters values, and calculation of the data rate, are conducted according to Table II and equations within \cite{Alzenad2018} for the clear sky scenario. As we see, the data rate ranges from an order of dozens to over a hundred gigabits per second (Gbps) for $P_t> 0$ dBm and H2G distance below 40 km. Practically, a powerful (resp. weak) H2G link with length 44 km would simultaneously support ITS services for 886 level 1, 147 level 2, 100 level 3, 84 level 4, and 63 level 5 (resp. 8 level 1, and 1 level 2) CAVs, respectively.
As the distance increases, the FSO data rate degrades rapidly. Hence, operating H2G with low transmit power or supporting a very remote HAPS-ITS node directly through a H2G link may not be feasible, in particular for level 3-5 CAVs. 

Fig. \ref{fig:FSO2} presents the data rate performance for FSO-based multi-hop backhaul links, where backhauling is modeled as a mesh communication network \cite{Jaafar2007} and successive HAPS-ITS nodes are separated horizontally by an 80 km distance (corresponding to the diameter of one HAPS-ITS node footprint). For the H2H link, we adopt the assumptions and equations of \cite{Fidler2010}, where FSO operates with wavelength 1550 nm and with the clear sky environment parameters as for Fig. \ref{fig:FSO1}. The data rate of the backhaul is dictated by the bottleneck link, defined as the minimum ratio of the link's capacity to its traffic load \cite{Jaafar2007}. We found that the data rate degrades as the number of H2H links increases due to a higher traffic load coming from the supported HAPS-ITS cells along the backhaul link. Nevertheless, multi-hop backhauling still achieves a better performance than the direct H2G link. For instance, the 121 km H2G link achieves a data rate of 3.5 Gbps (yellow curve in Fig \ref{fig:FSO1}), whereas an equivalent two-hop backhaul, with a 44 km H2G+80 km H2H, achieves data rate of 7.44 Gbps. Finally, we notice that the H2G link length impacts the bottleneck.     

In Fig. \ref{fig:mmWave}, we illustrate the data rate of the V2H and H2V/H2G links for different bandwidth values $B$, when conducted through mmWave. Our study is based on the 3GPP HAPS documentation, which was detailed in \cite[Subsection III-C.2]{Alfattani2021}. Typically, transmit powers of HAPS and CAV are 33 dBm (3GPP TR38.811) and 24 dBm (3GPP TR36.942), respectively. Also, we assume that antenna gains are 0 dBi except for the transmit gain of HAPS equal to 43.2 dBi \cite{Alfattani2021}. We set the operating frequency to $f=30$ GHz and consider rural environment conditions. Clearly, the mmWave H2G is less efficient than the FSO-based one, while the H2V achieves decent data rates with high bandwidth to support ITS services. For instance, at a distance of 40 km with $B=80$ MHz, the H2V link provides a downlink of 1.7 Gbps, which can support up to 4 level 5 CAVs simultaneously. Moreover, V2H performance is adequate for level 3-5 CAVs only with a large bandwidth (above 20 MHz) and for a relatively short distance (below 30 km).








\section{{Challenges and Future Directions}}

CAVs' safety must be always guaranteed. This condition is translated into a real-time and almost 100\% reliable data sensing and transmission requirements. A part of the solution consists on organizing and configuring a constellation of HAPS nodes into a redundant and real-time architecture. Moreover, current HAPS communication systems achieve reliability up to 99.9\%, which may be tolerable for level 1-3 CAVs, but still unsatisfying for level 4-5 CAVs that need reliability above 99.999\%. In that sense, new improvements and smarter antenna technologies must be developed.

Towards communication reliability, the use of hybrid FSO/RF may be leveraged for H2H and H2G communications. Although FSO and mmWave can be complimentary, a careful switching or combining design must be realized to bypass their sensitivity to weather conditions such as heavy rain, aerosols, and fog.
Moreover, 
orthogonal access techniques for H2H and H2G links can be leveraged to avoid interference. However, interference can be experienced by ground users due to inaccurate beamforming, overlapping at edge of HAPS cells, and dense HAPS constellations. Hence, more sophisticated access techniques and inter-HAPS coordination is needed to mitigate interference. Candidate techniques may include power control, antenna design, coordinated multi-point transmission, non-orthogonal multiple access, and m-MIMO.

Moreover, as work on HAPS design is ongoing, there is no clear consensus on what the lifetime of HAPS will be. Within the literature, energy consumption of HAPS related communication functionalities have been well investigated. However, since HAPS-ITS is envisioned to support advanced applications, requiring caching, computing, and imaging, payload type and energy consumption requirements need to be further discussed. To sustain long-term HAPS-ITS operations, more energy sources have to be explored, e.g., remote charging and nuclear energy.
Also, as HAPS nodes may experience failures, e.g., due to energy shortage, alternative approaches must be developed to continuously sustain the operations of CAVs. Potential solutions consist on the development of seamless redeployment strategies of HAPS nodes, and switching towards ground or satellite networks, if available, while accepting a potential degradation of services.
Finally, since LEO satellite mega-constellations are emerging to provide broadband Internet access across the globe (e.g., Starlink, OneWeb, Kuiper, and Telesat), the vertical integration of HAPS to LEO satellite networks for backhauling is necessary to achieve super connectivity \cite{kurt2020vision}. However, this integration is fragile as the HAPS network's performance heavily depends on the availability of satellite links, thus, making the HAPS system design more complex.


\section{Conclusion}
In this paper, we highlighted the potential of HAPS systems for supporting ITS services for trans-continental highways. Although ITS advancement have primarily been realized within vehicles, several crucial services require V2I communications, such as the assessment of road conditions, surveillance, platooning, and recovery in case of CAV failure. These services can be supported by a HAPS-ITS network located anywhere, but particularly in isolated areas where safety is a main concern. This was demonstrated through a case study, where a practical HAPS-ITS deployment design was presented and analyzed from a communications perspective.


 \section*{Acknowledgment}
\small{This work is funded by Huawei Canada and the National Science and Engineering Research Council of Canada. The authors thank Dr. Gamini Senarath, Huawei Canada, for valuable comments.}

\bibliographystyle{IEEEtran}  
\bibliography{references}

\section*{Biographies}
\small{

\noindent \textbf{Wael Jaafar} [SM] (waeljaafar@sce.carleton.ca) is an NSERC Postdoctoral Fellow in the SCE department of Carleton University, Canada. His research interests include aerial networks, 5G and beyond technologies, and machine learning.

\noindent \textbf{Halim Yanikomeroglu} [F] (halim@sce.carleton.ca) is a professor at Carleton University, Canada. His research interests cover 5G+ wireless networks. He is a Fellow of IEEE, the Engineering Institute of Canada (EIC), and the Canadian Academy of Engineering (CAE), and he is a Distinguished Speaker for IEEE Communications Society and IEEE VT Society.

}

\end{document}